\documentclass[aps,prb,twocolumn,showpacs]{revtex4}

\usepackage{graphicx}
\usepackage{amsmath}
\usepackage{amssymb}

\begin{document}

\title{Doping and momentum dependence of coupling strength in cuprate superconductors}

\author{Yingping Mou\footnotemark[1], Yiqun Liu\footnotemark[1], Shuning Tan, and Shiping Feng\footnotetext[1]{These authors contributed equally to this work}\footnote[2]{E-mail address: spfeng@bnu.edu.cn}}

\affiliation{Department of Physics, Beijing Normal University, Beijing 100875, China}

\begin{abstract}
Superconductivity is caused by the interaction between electrons by the exchange of bosonic excitations, however, this glue forming electron pairs is manifested itself by the coupling strength of the electrons to bosonic excitations. Here the doping and momentum dependence of the coupling strength of the electrons to spin excitations in cuprate superconductors is studied within the kinetic-energy-driven superconducting mechanism. The normal self-energy in the particle-hole channel and pairing self-energy in the particle-pariticle channel generated by the interaction between electrons by the exchange of spin excitation are employed to extract the coupling strengths of the electrons to spin excitations in the particle-hole and particle-particle channels, respectively. It is shown that below $T_{\rm c}$, both the coupling strengths in the particle-hole and particle-particle channels around the antinodes consist of two peaks, with a sharp low-energy peak located at 5 meV in the optimally doped regime, and a broad-band with a weak peak centered at 40 meV. In particular, this two-peak structure in the coupling strength in the particle-hole channel can persist into the normal-state, while the coupling strength in the particle-particle channel vanishes at the nodes. However, the positions of the peaks in the underdoped regime shift towards to higher energies with the increase of doping. More specifically, although the positions of the peaks move to lower energies from the antinode to the hot spot, the weights of the peaks decrease with the move of the momentum from the antinode to the hot spot, and fade away at the hot spots.
\end{abstract}


\maketitle

\section{Introduction}

Since the 1986 discovery of superconductivity in cuprate superconductors \cite{Bednorz86}, there has been an intense focus on the understanding of the essential physics of cuprate superconductors \cite{Anderson87}. This follows from a fact that the correlation between electrons in cuprate superconductors is so strong that when there is one electron on every copper atom site of their copper-oxide planes, a Mott insulator forms in which no electron motion is possible \cite{Anderson87,Phillips10,Kastner98}. By removing electrons, the electron motion is restored, and in particular, with the enough fraction of the removed electrons, superconductivity emerges \cite{Kastner98,Damascelli03,Zhou18}. At the temperature above the superconducting (SC) transition temperature $T_{\rm c}$, the system becomes a {\it strange-metal} \cite{Kastner98,Damascelli03,Zhou18,Comin16,Timusk99,Hufner08}, where a variety of the electronic orders are associated with various kinds of broken-symmetries. In particular, the resistivity is linear in temperature \cite{Timusk99,Hufner08,Batlogg94,Ando01}, the conductivity exhibits an anomalous power-law dependence on energy \cite{Timusk99,Hwang07}, and the electron Fermi surface (EFS) is broken up into the disconnected Fermi pockets around the nodal region \cite{Norman98,Yang08,Meng09,Yang11,Meng11,Ideta12,Kondo13}. All these anomalous properties arise from the strong interactions between the electrons mediated by the exchange of collective bosonic excitations \cite{Carbotte11,Choi18} that are most likely also responsible for the exceptionally high $T_{\rm c}$. In this case, it is crucial to elucidate the nature of the strong electron interaction for the understanding of the essential physics of cuprate superconductors.

In conventional superconductors, the Bardeen-Cooper-Schrieffer (BCS) theory has demonstrated that the coupling between the electrons and phonons drives the formation of the electron Cooper pairs responsible for superconductivity \cite{Bardeen57,Schrieffer64}. In particular, the strong coupling Eliashberg equation for the determination of $T_{\rm c}$ depends upon the coupling strength of the electrons to phonons \cite{Eliashberg60,McMillan65,Carbotte90,Basov05}, which is manifested itself by the electron-phonon spectral density (then the Eliashberg function) $\alpha^{2}F_{\rm e-p}(\omega)$. On the other hand, in a remarkable analogy to conventional superconductors, the pairing of electrons in cuprate superconductors also occurs at $T_{\rm c}$, creating an energy gap in the quasiparticle excitation spectrum \cite{Kastner98,Damascelli03,Zhou18}. The BCS theory and the related Eliashberg formalism are not specific to a phonon-mediated interaction, other collective bosonic excitations can also serve as the pairing glue \cite{Miller11,Monthoux07}. However, what type of the collective bosonic excitation that acts like a bosonic glue to hold the electron Cooper pairs together in cuprate superconductors still is disputed \cite{Carbotte11,Choi18}. Moreover, in a stark contrast to the conventional superconductors, the coupling strength in cuprate superconductors has an important momentum dependence \cite{Tsuei00,Hardy93,Wollman93,Tsuei94}, so that one should like to understand how both momentum {\bf k} and energy $\omega$ regions contribute to $T_{\rm c}$. Experimentally, from the angle-resolved photoemission spectroscopy (ARPES) data of cuprate superconductors measured below and above $T_{\rm c}$, with the analysis of the normal Green's function in the particle-hole channel and pairing Green's function in the particle-particle channel, the doping and momentum dependence of the normal self-energy in the particle-hole channel and pairing self-energy in the particle-particle channel have been successfully extracted \cite{Zhang08,Zhang12,Bok10,Bok16}, respectively. In particular, although the Eliashberg functions can not be measured directly from the ARPES experiments, by a direct inversion of the Eliashberg function, the doping and momentum dependence of the normal Eliashberg function $\alpha^{2}F^{\rm (n)}_{\rm e-b}({\bf k},\omega)$ in the particle-hole channel and pairing Eliashberg function $\alpha^{2}F^{\rm (p) }_{\rm e-b}({\bf k},\omega)$ in the particle-particle channel have been deduced from the corresponding normal and pairing self-energies based on the maximum entropy method \cite{Bok10,Bok16,Yun11,Schachinger08}, where both the normal and pairing Eliashberg functions $\alpha^{2}F^{\rm (n)}_{\rm e-b}({\bf k},\omega)$ and $\alpha^{2}F^{\rm (p)}_{\rm e-b}({\bf k},\omega)$ show two-peaks \cite{Yun11} at around 15 meV and 50 meV, followed by a flat featureless region that extends to high energy, and a cut-off at high energy depends on the location of the momentum cuts. Furthermore, both the peaks become enhanced as the temperature is lowered or the positions of these peaks move away from the node to the antinode \cite{Bok10,Bok16,Yun11,Schachinger08}. These experiments and analysis in cuprate superconductors are an extension \cite{Vekhter03} of the tunneling experiments and analysis with which it was definitively established that the Cooper pairing in conventional superconductors is through exchange of phonons \cite{Eliashberg60,McMillan65,Carbotte90,Basov05}. Theoretically, the numerical simulations of the generalized Eliashberg equation with a d-wave SC gap have been provided to help in the understanding of how the structure of collective bosonic excitations encoded in the coupling strength present itself in the normal and pairing self-energies \cite{Scalapino12,Schachinger09,Kyung09,Kancharla08,Haule07}. In particular, it has been shown \cite{Maier08} based on the $t$-$J$ model that the dominant contribution to the electron Cooper pairing in cuprate superconductors is associated with the coupling strength of the electrons to spin excitations $\alpha^{2}F_{\rm e-s}({\bf k},\omega)$. These deduced normal and pairing Eliashberg functions from the ARPES measurements \cite{Bok10,Bok16,Yun11,Schachinger08} and the numerical simulations of the generalized Eliashberg equations \cite{Scalapino12,Schachinger09,Kyung09,Kancharla08,Haule07,Maier08} therefore provide critical information on the examination of various microscopic SC mechanisms. In this paper, we study how the coupling strength of the electrons to spin excitations in cuprate superconductors evolves with momentum, energy, and doping. Within the framework of the kinetic-energy driven SC mechanism \cite{Feng0306,Feng12,Feng15}, both the normal and pairing self-energies are generated by the strong interaction between electrons by the exchange of spin excitations. We then employ these normal and pairing self-energies to extract the normal and pairing Eliashberg functions $\alpha^{2}F^{\rm (n) }_{\rm e-s}({\bf k},\omega)$ and $\alpha^{2}F^{\rm (p)}_{\rm e-s}({\bf k},\omega)$, respectively, and the obtained results show that below $T_{\rm c}$, both the coupling strengths in the particle-hole and particle-particle channels around the antinodes exhibit a two-peak structure, where a sharp low-energy peak develops at around 5 meV in the optimally doped regime, followed by a broad band with a weak peak centered at around 40 meV, in qualitative agreement with the results deduced from the ARPES experimental observations on cuprate superconductors \cite{Yun11}. In particular, this two-peak structure in the coupling strength in the particle-hole channel can persist into the normal-state. However, these coupling strengths are doping dependent, where the positions of the peaks in the coupling strengths in the underdoped regime shift towards to higher energies with the increase of doping. Furthermore, these coupling strengths also have a striking momentum dependence. Although the positions of the peaks in the coupling strengths move to lower energies from the antinode to the hot spot on EFS, the weights of the peaks decrease smoothly with the move of the momentum from the antinode to the hot spot, and fade away at the hot spots.

This paper is organized as follows. The basic formalism of the coupling strength of the electrons to spin excitations in cuprate superconductors is presented in Sec. \ref{Coupling-strength}, while the quantitative characteristics of the coupling strength are discussed in Sec. \ref{Eliashberg-function}, where we show that as a consequence of the d-wave type symmetry of the SC gap, the coupling strength in the particle-particle channel vanishes at the nodes. Finally, we give a summary in Sec. \ref{conclusions}.

\section{Coupling strength of electrons to spin excitations}\label{Coupling-strength}

Experimentally, the normal and pairing self-energies can be extracted from ARPES experiments without a specific theory, and the only ingredient that needs to extract the normal and pairing self-energies is the quasiparticle spectral density observed by ARPES experiments. However, to deduce the normal and pairing Eliashberg functions from the extracted normal and pairing self-energies, respectively, one needs a microscopic SC theory relating them \cite{Carbotte11,Choi18}. This is why the qualitative agreement between the experimental result observed from the tunneling spectra and the deduced coupling strength of the electrons to phonons from the Eliashberg equation firmly establishes the electron-phonon mechanism \cite{Eliashberg60,McMillan65,Carbotte90,Basov05}. In this case, to examine a microscopic SC theory, it should be to compare the normal and pairing Eliashberg functions rather than the normal and pairing self-energies \cite{Carbotte11,Choi18}. Our following work for the discussions of the coupling strength of the electrons to spin excitations in cuprate superconductors builds on the framework of the kinetic-energy driven superconductivity \cite{Feng0306,Feng12,Feng15}, which was developed early based on the $t$-$J$ model in the charge-spin separation fermion-spin representation. In this kinetic-energy driven SC mechanism, cuprate superconductors involve the d-wave type charge-carrier pairs bound together by the interaction between charge carriers and spins directly from the kinetic-energy of the $t$-$J$ model by the exchange of {\it spin excitations}, then the d-wave type electron Cooper pairs originating from the d-wave type charge-carrier pairing state are due to the charge-spin recombination \cite{Feng15a}, and they condense to the SC ground-state. In particular, in our previous works \cite{Feng15a}, the normal self-energy $\Sigma_{1}({\bf k},\omega)$ and pairing self-energy $\Sigma_{2}({\bf k},\omega)$ have been evaluated in terms of the full charge-spin recombination, and can be written explicitly as,
\begin{eqnarray}
\Sigma_{1}({\bf k},\omega)&=&\int^{\infty}_{-\infty}{{\rm d}\omega'\over 2\pi}\int^{\infty}_{-\infty}{{\rm d}\omega''\over 2\pi}{n_{\rm B}(\omega'')+ n_{\rm{F}}(\omega')\over\omega''-\omega'+\omega}\nonumber\\
&\times&{1\over N}\sum_{\bf p}A({\bf p},\omega'){\bar K}_{\rm e-s}({\bf k},{\bf p},\omega''), ~~~ \label{NSE} \\
\Sigma_{2}({\bf k},\omega)&=&\int^{\infty}_{-\infty}{{\rm d}\omega'\over 2\pi}\int^{\infty}_{-\infty}{{\rm d}\omega''\over 2\pi}{n_{\rm B}(\omega'')+ n_{\rm{F}}(\omega')\over\omega''-\omega'+\omega}\nonumber\\
&\times& {1\over N}\sum_{\bf p}A_{\Im^{\dagger}}({\bf p},\omega'){\bar K}_{\rm e-s}({\bf k}, {\bf p}, \omega''), ~~~\label{PSE}
\end{eqnarray}
respectively, where $n_{\rm F}(\omega)$ and $n_{\rm B}(\omega)$ are the fermion and boson distribution functions, respectively, the normal spectral function $A({\bf k},\omega)=-2{\rm Im}G({\bf k},\omega)$ and the pairing spectral function $A_{\Im^{\dagger}}({\bf k},\omega)=-2{\rm Im} \Im^{\dagger}({\bf k},\omega)$ are related directly to the imaginary parts of the normal and pairing Green's functions of the $t$-$J$ model $G({\bf k},\omega)$ and $\Im^{\dagger}({\bf k},\omega)$, respectively, with the normal Green's function $G({\bf k},\omega)$ and pairing Green's function $\Im^{\dagger}({\bf k},\omega)$ that have been given explicitly in Ref. \cite{Feng15a}, while the kernel function ${\bar K}_{\rm e-s}({\bf k}, {\bf p},\omega)$ describes the nature of the spin excitations, and can be expressed explicitly as,
\begin{eqnarray}\label{kernel}
{\bar K}_{\rm e-s}({\bf k},{\bf p},\omega)={2\over N}\sum_{\bf q}\Lambda^{2}_{{\bf p}+{\bf q}}{\rm Im}\Pi({\bf k},{\bf p},{\bf q},\omega).
\end{eqnarray}
where $\Lambda_{{\bf k}}=Zt\gamma_{\bf k}-Zt'\gamma_{\bf k}'$, with the electron nearest-neighbor (NN) and next NN hopping integrals $t$ and $t'$ in the $t$-$J$ model, respectively, the number of the NN or next NN sites on a square lattice $Z$, $\gamma_{{\bf k}}=({\rm cos}k_{x}+{\rm cos}k_{y})/2$, $\gamma_{\bf k}'= {\rm cos} k_{x}{\rm cos}k_{y}$, and ${\rm Im}\Pi({\bf k},{\bf p},{\bf q},\omega)$ is the imaginary part of the spin bubble $\Pi({\bf k},{\bf p},{\bf q},\omega)$, with this spin bubble $\Pi({\bf k}, {\bf p},{\bf q},\omega)$ that is a convolution of two spin Green's functions, and has been also given explicitly in Ref. \cite{Feng15a}. Within the framework of the kinetic-energy driven superconductivity, the normal self-energy $\Sigma_{1}({\bf k},\omega)$ of cuprate superconductors in the particle-hole channel in Eq. (\ref{NSE}), which describes the single-particle coherence, and the pairing self-energy $\Sigma_{2}({\bf k},\omega)$ in the particle-particle channel in Eq. (\ref{PSE}), which is corresponding to the energy for breaking a Cooper pair of the electrons and creating two excited states, are generated by the same interaction of electrons with spin excitations. These normal and pairing self-energies exhibit a strong momentum anisotropy in both the real and imaginary parts, and are essential to understand the electronic state properties and the related quasiparticle dynamics of cuprate superconductors in the SC-state.

With the help of the above normal and pairing self-energies in Eqs. (\ref{NSE}) and (\ref{PSE}), the coupling strength of the electrons to spin excitations in the particle-hole channel (then the normal Eliashberg function) can be evaluated by the average over all momenta weighted by the normal spectral function as \cite{Choi18,Eliashberg60,McMillan65,Carbotte90,Basov05,Mahan81},
\begin{eqnarray}\label{NEF}
\alpha^{2}F^{\rm (n)}_{\rm e-s}({\bf k},\omega)&=&{1\over N}\sum_{\bf p}A({\bf p},\omega){\bar K}_{\rm e-s}({\bf k},{\bf p},\omega),
\end{eqnarray}
which therefore is related closely with the normal self-energy in Eq. (\ref{NSE}), while the coupling strength of the electrons to spin excitations in the particle-particle channel (then the pairing Eliashberg function) can be evaluated by the average over all momenta weighted by the pairing spectral function as \cite{Choi18,Eliashberg60,McMillan65,Carbotte90,Basov05,Mahan81},
\begin{eqnarray}\label{PEF}
\alpha^{2}F^{\rm (p)}_{\rm e-s}({\bf k},\omega)&=&{1\over N}\sum_{\bf p}A_{\Im^{\dagger}}({\bf p},\omega){\bar K}_{\rm e-s}({\bf k},{\bf p},\omega),
\end{eqnarray}
which is associated directly with the pairing self-energy in Eq. (\ref{PSE}). It is thus shown that the structures of the normal and pairing self-energies in Eqs. (\ref{NSE}) and (\ref{PSE}) are reflected in the coupling strengths of the electrons to spin excitations in the particle-hole and particle-particle channels in Eqs. (\ref{NEF}) and (\ref{PEF}), respectively.

\section{Quantitative characteristics} \label{Eliashberg-function}

\begin{figure}[t!]
\centering
\includegraphics[scale=1.25]{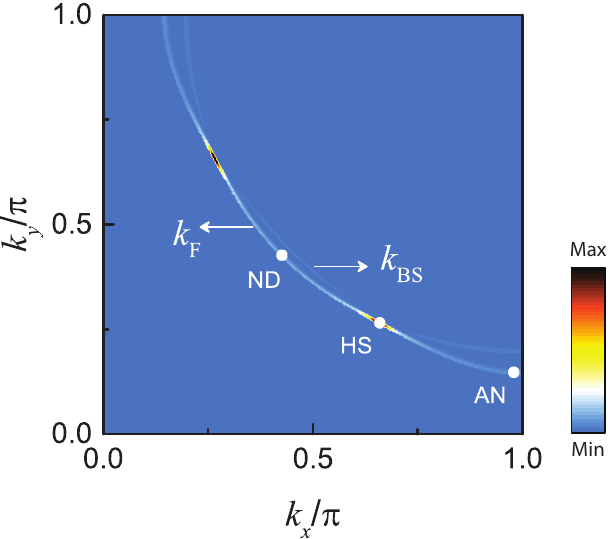}
\caption{(Color online) The map of the normal spectral function in the $[k_{x},k_{y}]$ plane at $\delta=0.15$ with $T=0.002J$ for $t/J=2.5$ and $t'/t=0.3$, where AN, HS, and ND denote the antinode, hot spot, and node, respectively. \label{spectral-map}}
\end{figure}

For an interacting electron system, everything happens at EFS \cite{Mahan81}. This is why EFS plays a crucial role in the understanding of the physical properties of the interacting electron system. For a convenience in the following discussions, the underlying EFS of cuprate superconductors at doping $\delta=0.15$ is {\it replotted} in Fig. \ref{spectral-map}, which is a map \cite{Gao18a} of the intensity of the normal spectral function $A({\bf k},\omega=0)$ at zero binding energy $\omega=0$ with temperature $T=0.002J$ for parameters $t/J=2.5$ and $t'/t=0.3$. AN, HS, and ND in Fig. \ref{spectral-map} denote the antinode, hot spot, and node on EFS, respectively, where the hot spot is determined by the highest peak heights on EFS \cite{Gao18a}. As we \cite{Gao18a} have shown in the previous discussions, the most remarkable features in Fig. \ref{spectral-map} can be summarized as: (i) the Fermi pocket emerges around the nodal region due to the EFS reconstruction \cite{Norman98,Yang08,Meng09,Yang11,Meng11,Ideta12,Kondo13}, where the disconnected segment at the contour ${\bf k}_{\rm F}$ is referred to the Fermi arc, while the other at the contour ${\bf k}_{\rm BS}$ is defined as the back side of the Fermi pocket. This EFS reconstruction with the anisotropic distribution of the quasiparticle spectral weight also indicates that the information around the nodes, hot spots, and antinodes contain the essentials of the whole low-energy quasiparticle excitations in cuprate superconductors; (ii) however, the highest intensity points do not locate at the node places, but sit exactly at the tips of the Fermi arcs \cite{Sassa11,Shi08,Chatterjee06}, which in this case coincide with the hot spots on EFS; (iii) these eight hot spots connected by the scattering wave vectors ${\bf q}_{i}$ construct an {\it octet} scattering model \cite{Chatterjee06,Gao18b,Pan01,Kohsaka07,Hanaguri07,Kohsaka08,Hanaguri09,Kondo09,Vishik09}, and therefore contribute effectively to the quasiparticle scattering process. More specifically, this EFS instability drives charge order \cite{Comin16,Gao18a,Comin14,Wu11,Campi15,Comin15,Hinton16,Feng16}, with the charge-order wave vector that matches well with the wave vector connecting the two hot spots on the straight Fermi arcs. Furthermore, it should be emphasized that all these typical features occurred at the case of zero binding energy are almost the same as the case for finite binding energies \cite {Gao18b}.

\begin{figure}[t!]
\centering
\includegraphics[scale=1.05]{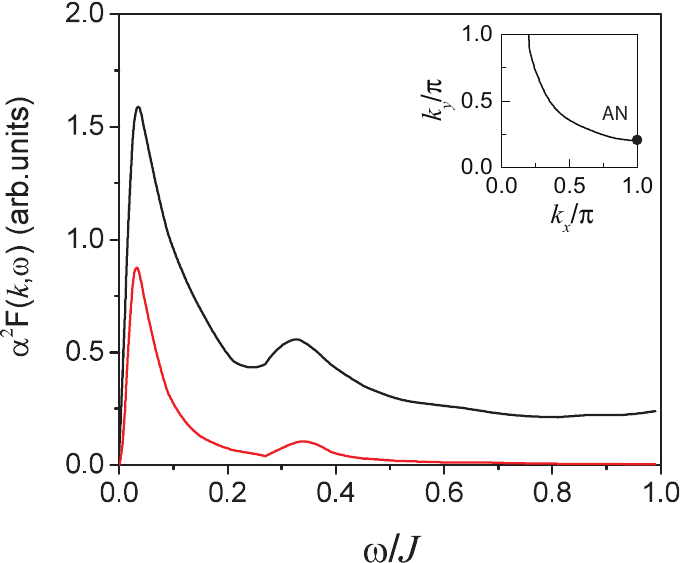}
\caption{(Color online) The normal (black line) and pairing (red line) Eliashberg functions as a function of binding-energy at the antinode in $\delta=0.15$ with $T=0.002J$ for $t/J=2.5$ and $t'/t=0.3$.  \label{coupling-strength-AN}}
\end{figure}

We are now ready to discuss the quantitative characteristics of the doping and momentum dependence of the coupling strength of the electrons to spin excitations in cuprate superconductors based on the kinetic-energy driven superconductivity. In Fig. \ref{coupling-strength-AN}, we plot the normal (black line) and pairing (red line) Eliashberg functions $\alpha^{2}F^{\rm (n)}_{\rm e-s}({\bf k},\omega)$ and $\alpha^{2}F^{\rm (p)}_{\rm e-s}({\bf k},\omega)$, respectively, as a function of binding-energy at the antinode in $\delta=0.15$ with $T=0.002J$ for $t/J=2.5$ and $t'/t=0.3$, where for the temperature below $T_{\rm c}$, both $\alpha^{2}F^{\rm (n)}_{\rm e-s}({\bf k},\omega)$ and $\alpha^{2}F^{\rm (p)}_{\rm e-s}({\bf k}, \omega)$ extends over a broad energy range. In particular, both $\alpha^{2}F^{\rm (n)}_{\rm e-s}({\bf k},\omega)$ and $\alpha^{2}F^{\rm (p)}_{\rm e-s} ({\bf k},\omega)$ contain a sharp low-energy peak located at around $\omega_{\rm LP}\sim 0.04J$ and a broad band, followed by a flat featureless region extending to high energy. Moreover, this broad band shows a weak peak centered at around $\omega_{\rm BP}\sim 0.34J$. However, the weights of the peaks in $\alpha^{2}F^{\rm (p)}_{\rm e-s}({\bf k},\omega)$ are smaller than that in $\alpha^{2}F^{\rm (n)}_{\rm e-s}({\bf k},\omega)$. The interaction of the electrons with spin excitations is therefore characterized by these peaks in the coupling strength. Using a reasonably estimative value of $J\sim 120$ meV in cuprate superconductors \cite{Kastner98,Fujita12}, the positions of the sharp low-energy peak and the broad peak are located at around $\omega_{\rm LP}\approx 5$ meV and $\omega_{\rm BP}\approx 40$ meV, respectively, which are not too far from the corresponding sharp low-energy peak $\omega_{\rm LP} \approx 15$ meV and the broad peak $\omega_{\rm BP}\approx 50$ meV in the coupling strengths deduced from the ARPES experimental data \cite{Yun11} of the cuprate superconductor Bi$_{2}$Sr$_{2}$CaCu$_{2}$O$_{8+\delta}$.

\begin{figure}[t!]
\centering
\includegraphics[scale=1.05]{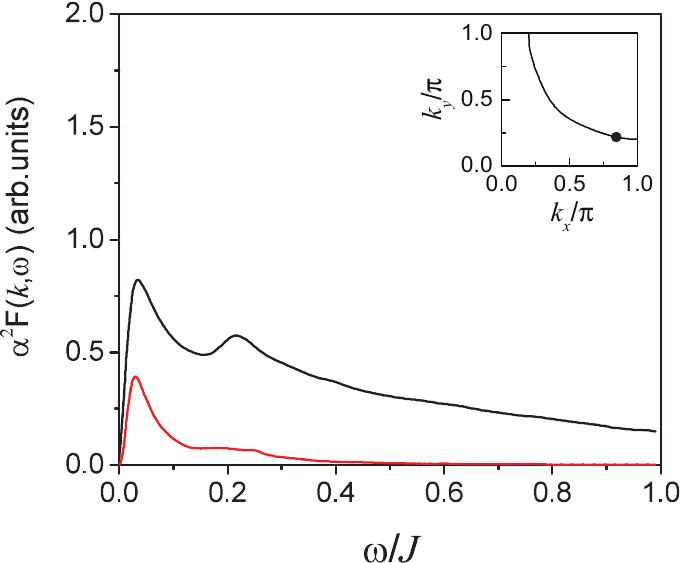}
\caption{(Color online) The normal (black line) and pairing (red line) Eliashberg function as a function of binding-energy away from the antinode at $\delta=0.15$ with $T=0.002J$ for $t/J=2.5$ and $t'/t=0.3$.  \label{coupling-strength-ND}}
\end{figure}

However, the anisotropic distribution of the quasiparticle spectral weight along EFS shown in Fig. \ref{spectral-map} due to the EFS reconstruction also indicates that in contrast to conventional superconductors, the coupling strength of the electrons to spin excitations in cuprate superconductors is strongly momentum dependent. To show this point clearly, we plot the normal (black line) and pairing (red line) Eliashberg functions, respectively, away from the antinode as a function of binding-energy at $\delta=0.15$ with $T=0.002J$ for $t/J=2.5$ and $t'/t=0.3$ in Fig. \ref{coupling-strength-ND}. Comparing it with the result in Fig. \ref{coupling-strength-AN} for the same set of parameters except for the momentum ${\bf k}_{\rm F}$ away from the antinode, we therefore find when the momentum ${\bf k}_{\rm F}$ moves from the antinode to the hot spot, the weights of the peaks in the coupling strengths decrease, while the positions of the peaks shift appreciably towards to lower energies, in qualitative agreement with the results deduced from the ARPES experimental data of cuprate superconductors \cite{Bok10,Bok16,Yun11,Schachinger08}. More surprisedly, we find during the calculations that the coupling strengths in both the particle-hole and particle-particle channels have an unusual momentum dependence around the hot spot region with the anomalously small value at the hot spots, where $\alpha^{2}F^{\rm (n)}_{\rm e-s}({\bf k}_{\rm HS},\omega)\sim 0$ and $\alpha^{2}F^{\rm (p)}_{\rm e-s}({\bf k}_{\rm HS},\omega)\sim 0$. Furthermore, we also find that the coupling strength $\alpha^{2}F^{\rm (p)}_{\rm e-s}({\bf k}_{\rm ND},\omega)$ in the particle-particle channel vanishes at the nodes, i.e., $\alpha^{2}F^{\rm (p)}_{\rm e-s}({\bf k}_{\rm ND},\omega)=0$.

\begin{figure}[t!]
\centering
\includegraphics[scale=1.05]{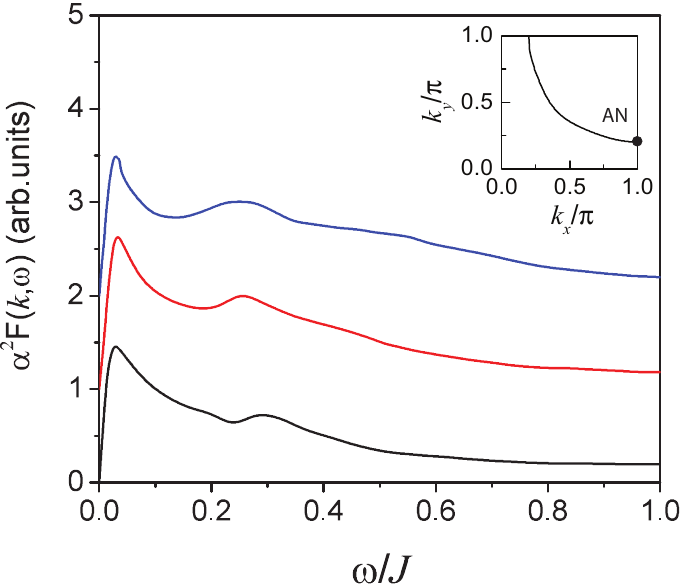}
\caption{(Color online) The normal Eliashberg function as a function of binding-energy at the antinode in $\delta=0.06$ (blue line), $\delta=0.09$ (red line), and $\delta=0.12$ (black line) with $T=0.002J$ for $t/J=2.5$ and $t'/t=0.3$. \label{coupling-strength-doping}}
\end{figure}

As a natural consequence of the doped Mott insulators, the coupling strength of the electrons to spin excitations in cuprate superconductors is also doping dependent. For a better understanding of the evolution of the coupling strength of the electrons to spin excitations with doping, we have made a series of calculations for the Eliashberg functions at different doping concentration, and the results of the normal Eliashberg function $\alpha^{2}F^{\rm (n)}_{\rm e-s}({\bf k},\omega)$ as a function of binding-energy at the antinode in $\delta=0.06$ (blue line), $\delta=0.09$ (red line), and $\delta=0.12$ (black line) with $T=0.002J$ for $t/J=2.5$ and $t'/t=0.3$ are plotted Fig. \ref{coupling-strength-doping}, where in the underdoped regime, although the positions of the sharp low-energy peaks are insensitive to the doping concentration, the positions of the weak peaks at the broad band move smoothly towards to the higher energies with the increase of doping, also in qualitative agreement with the results deduced from the ARPES experimental data of cuprate superconductors \cite{Bok10,Bok16,Yun11,Schachinger08}.

\begin{figure}[t!]
\centering
\includegraphics[scale=1.05]{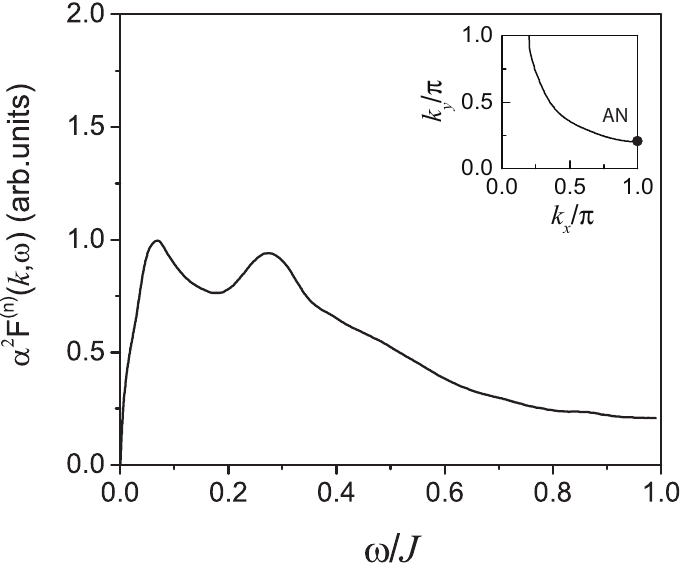}
\caption{The normal Eliashberg function as a function of energy at the antinode in $\delta=0.09$ with $T=0.06J$ above $T_{\rm c}$ for $t/J=2.5$ and $t'/t=0.3$. \label{coupling-strength-temperature}}
\end{figure}

These coupling strengths of the electrons to spin excitations in both the particle-hole and particle-particle channels are also temperature dependent. In particular, for the temperatures above $T_{\rm c}$, the SC gap $\bar{\Delta}_{\rm s}({\bf k},\omega)=\Sigma_{2}({\bf k},\omega)=0$, indicating that the coupling strength of the electrons to spin excitations in the particle-particle channel in Eq. (\ref{PEF}) disappears for $T\geq T_{\rm c}$. However, the coupling strength of the electrons to spin excitations in the particle-hole channel evolves smoothly into the normal-state as the temperature is raised above $T_{\rm c}$. To further reveal the unusual behavior of the coupling strength of the electrons to spin excitations, we have also performed a series of calculations for the normal Eliashberg function with different temperatures, and the result of $\alpha^{2}F^{\rm (n)}_{\rm e-s}({\bf k},\omega)$ as a function of binding-energy at the antinode in $\delta=0.09$ with $T=0.06J$ for $t/J=2.5$ and $t'/t=0.3$ is plotted Fig. \ref{coupling-strength-temperature}. Within the framework of the kinetic-energy driven superconductivity \cite{Feng15a}, the calculated $T_{\rm c}\sim 0.05J$ at the doping $\delta=0.09$. Obviously, the two-peak structure of the coupling strength of the electron to spin excitations appeared in the SC-state persists into the normal-state. On comparison with the result (red line) in Fig. \ref{coupling-strength-doping} for the same set of parameters except for $T=0.06J$, we see that although the weights of the peaks are severely suppressed with the increase of temperatures, the positions of these peaks in the normal-state do not change much from the corresponding SC-state. These results are qualitatively consistent with the results deduced from the ARPES experimental data of cuprate superconductors \cite{Bok10,Bok16,Yun11,Schachinger08}.

\begin{figure*}[t!]
\centering
\includegraphics[scale=1.35]{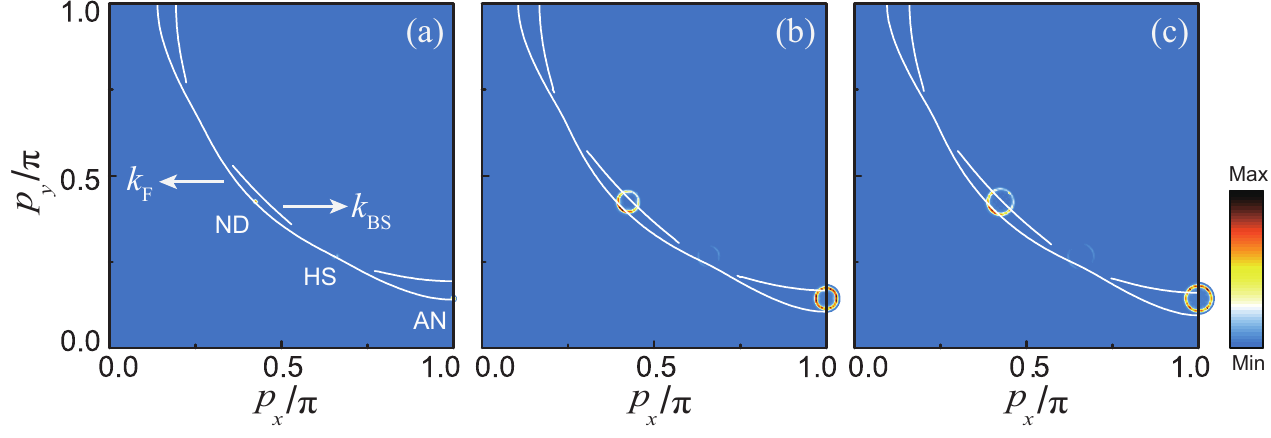}
\caption{(Color online) The maps of three typical kernel functions, ${\bar K}_{\rm e-s}({\bf k}_{\rm AN},{\bf p},\omega)$ at the antinode ${\bf k}_{\rm AN}$, ${\bar K}_{\rm e-s}({\bf k}_{\rm HS},{\bf p},\omega)$ at the hot spot ${\bf k}_{\rm HS}$, and ${\bar K}_{\rm e-s}({\bf k}_{\rm ND},{\bf p},\omega)$ at the node ${\bf k}_{\rm ND}$, in a $[p_{x},p_{y}]$ plane for the binding-energy (a) $\omega=0.04J$, (b) $\omega=0.34J$, and (c) $\omega=0.40J$ in $T=0.002J$ and $\delta=0.15$ with $t/J=2.5$ and $t'/t=0.3$. \label{kernel-function}}
\end{figure*}

Now we give some physical interpretations to the above obtained results. As seen from Eqs. (\ref{NEF}) and (\ref{PEF}), the coupling strengths of the electrons to spin excitations are determined by a product of the quasiparticle spectral function and the kernel function ${\bar K}_{\rm e-s}({\bf k}, {\bf p},\omega)$. In other words, for a given momentum ${\bf k}$, the dominant contribution to the coupling strength at the binding-energy is if and only if the spectral region characteristic of the quasiparticles overlaps with the spectral region of the kernel function characteristic of the spin excitations at that binding-energy.

As the result shown in Fig. \ref{spectral-map}, the single continuous contour in momentum space with a uniform distribution of the spectral weight of the quasiparticle excitations in the case of the absence of the coupling between the electrons and spin excitations has been split into two contours ${\bf k}_{\rm F}$ and ${\bf k}_{\rm BS}$, respectively, in the presence of the coupling between the electrons and spin excitations to form the Fermi pockets with the anisotropic distribution of the spectral weight of the quasiparticle excitations.

On the one hand, for a given momentum ${\bf k}$, the kernel function ${\bar K}_{\rm e-s}({\bf k},{\bf p},\omega)$ also exhibits a remarkable evolution with momentum ${\bf p}$ and binding-energy $\omega$ except for zero binding-energy $\omega=0$, where the kernel function ${\bar K}_{\rm e-s}({\bf k},{\bf p},\omega) |_{\omega=0}=0$. To see the unusual momentum dependence of the kernel function clearly, we plot the maps of three typical kernel functions, ${\bar K}_{\rm e-s}({\bf k}_{\rm AN},{\bf p},\omega)$ at the antinode ${\bf k}_{\rm AN}$, ${\bar K}_{\rm e-s} ({\bf k}_{\rm HS}, {\bf p},\omega)$ at the hot spot ${\bf k}_{\rm HS}$, and ${\bar K}_{\rm e-s}({\bf k}_{\rm ND}, {\bf p},\omega)$ at the node ${\bf k}_{\rm ND}$, in a $[p_{x},p_{y}]$ plane for the binding-energy (a) $\omega=0.04J$, (b) $\omega=0.34J$, and (c) $\omega=0.40J$ with $T=0.002J$ and $\delta=0.15$ for $t/J=2.5$ and $t'/t=0.3$ in Fig. \ref{kernel-function}. At low binding-energies (see Fig. \ref{kernel-function}a), the weight of the momentum ${\bf p}$ dependence of the kernel function ${\bar K}_{\rm e-s}({\bf k}_{\bf F}, {\bf p},\omega)$ at a given ${\bf k}_{\bf F}$ point converges on the corresponding ${\bf p}={\bf k}_{\bf F}$ point, i.e., ${\bar K}_{\rm e-s}({\bf k}_{\bf F},{\bf p}={\bf k}_{\bf F},\omega)\neq 0$ for ${\bf p}={\bf k}_{\bf F} $, and otherwise ${\bar K}_{\rm e-s}({\bf k}_{\bf F},{\bf p},\omega)=0$. In particular, the weight of ${\bar K}_{\rm e-s}({\bf k}_{\bf AN} ,{\bf p}={\bf k}_{\bf AN},\omega)$ exhibits a largest value at around the antinode ${\bf k}_{\rm AN}$. This strong overlap between the weights of the normal spectral function $A({\bf p}={\bf k}_{\bf AN},\omega)$ and the kernel function ${\bar K}_{\rm e-s}({\bf k}_{\bf AN},{\bf p}={\bf k}_{\bf AN}, \omega)$ at around the antinode therefore leads to the appearance of the sharp low-energy peak of the coupling strength in the particle-hole channel at the low binding-energy $\omega=0.04J$ for the doping $\delta=0.15$ as shown in Fig. \ref{coupling-strength-AN} (black line). However, the weight of ${\bar K}_{\rm e-s}({\bf k}_{\bf F},{\bf p}={\bf k}_{\bf F},\omega)$ on EFS is angular dependent, i.e., when the momentum ${\bf k}_{\bf F}$ moves away from the antinode and towards to the hot spot, the weight of ${\bar K}_{\rm e-s}({\bf k}_{\bf F},{\bf p}={\bf k}_{\bf F},\omega)$ on EFS decreases, which induces a reduction of the weight of the low-energy peak in the coupling strength as shown in Fig. \ref{coupling-strength-ND} (black line). In particular, the weight of ${\bar K}_{\rm e-s} ({\bf k}_{\bf HS},{\bf p}={\bf k}_{\bf HS},\omega)$ almost vanishes at the hot spots, where ${\bar K}_{\rm e-s}({\bf k}_{\rm HS},{\bf p}={\bf k}_{\bf HS}, \omega)\sim 0$, leading to an absence of the coupling of the electrons to spin excitations at the hot spots $\alpha^{2}F^{\rm (n)}_{\rm e-s}({\bf k}_{\rm HS},\omega)\sim 0$. However, the weight of ${\bar K}_{\rm e-s} ({\bf k}_{\rm F},{\bf p} ={\bf k}_{\rm F},\omega)$ on EFS smoothly increases when the momentum ${\bf k}_{\bf F}$ moves away from the hot spot and towards to the node, which generates that the coupling strength gradually develops again with the move of the momentum ${\bf k}_{\bf F}$ from the hot spot to the node, and then the coupling strength of the electrons to spin excitations has a modest value at around the node.

On the other hand, as shown in Figs. \ref{kernel-function}b and \ref{kernel-function}c, for a given momentum ${\bf k}_{\bf F}$, the area occupied by the weight of the kernel function ${\bar K}_{\rm e-s}({\bf k}_{\bf F},{\bf p},\omega)$ increases with the increase of binding-energy $\omega$. In other words, for a given momentum ${\bf k}_{\bf F}$, the weight of ${\bar K}_{\rm e-s}({\bf k}_{\bf F},{\bf p},\omega)$ that converges on ${\bf p} ={\bf k}_{\bf F}$ at low binding-energies lies on a circle at higher binding-energies, with the circle that increases in radius with binding-energy, while the distribution of the weight of ${\bar K}_{\rm e-s}({\bf k}_{\bf F},{\bf p},\omega)$ on this circle is rather isotropic. In particular, at the higher binding-energy $\omega=0.34J$, the part of the weight of ${\bar K}_{\rm e-s}({\bf k}_{\bf AN},{\bf p}={\bf k}_{\rm BS},\omega)$ on the circle overlaps with the spectral weight of the normal spectral function $A({\bf p}={\bf k}_{\rm BS},\omega)$ in the constant energy contour ${\bf k}_{\rm BS}$ at around the antinode for the doping $\delta=0.15$ as shown in Fig. \ref{kernel-function}b, which therefore leads to the emergence of the broad peak in the coupling strength of the electrons to spin excitations centered at around $\omega=0.34J$ as shown in Fig. \ref{coupling-strength-AN}. However, for the binding-energies $\omega>0.34J$ in the doping $\delta=0.15$, there is no overlap between the weight of ${\bar K}_{\rm e-s}({\bf k}_{\bf F},{\bf p},\omega)$ and the normal spectral function $A({\bf p},\omega)$ at both the constant energy contours ${\bf k}_{\rm F}$ and ${\bf k}_{\rm BS}$ as shown in Fig. \ref{kernel-function}c, and then a flat featureless region extending to higher energy appears in the coupling strength as shown in Fig. \ref{coupling-strength-AN}.

The essential physics of the coupling strength of the electrons to spin excitations in the particle-particle channel is exactly the same as the above discussion for the coupling strength in the particle-hole channel except for the nodal regime. Within the framework of the kinetic energy driven superconductivity, the SC-state of cuprate superconductors has a strong anisotropic d-wave type symmetry with the SC gap on EFS vanishing along the main diagonals of the Brillouin zone \cite{Feng15a}, i.e., $\bar{\Delta}_{\rm s}({\bf k},\omega)|_{{\bf k}={\bf k}_{\rm ND}}=\Sigma_{2}({\bf k}, \omega)|_{{\bf k}={\bf k}_{\rm ND}}=0$ at the nodes, which therefore leads to that the coupling strength of the electrons to spin excitations in the particle-particle channel vanishes at the nodes, i.e., $\alpha^{2}F^{\rm (p)}_{\rm e-s}({\bf k}, \omega)|_{{\bf k}={\bf k}_{\rm ND}}=0$.

The above discussions therefore show that the remarkable feature of the anisotropy of EFS due to the redistribution of the quasiparticle spectral weight on the constant energy contours ${\bf k}_{\rm F}$ and ${\bf k}_{\rm BS}$, and the unusual momentum and energy dependence of the kernel function ${\bar K}_{\rm e-s}({\bf k}_{\bf F},{\bf p},\omega)$ generate the two-peak structure in the coupling strength of the electrons to spin excitations along with EFS except for the hot spots: the sharp low-energy peak in the coupling strength associated with the overlap of the weights of the quasiparticle spectral function and kernel function ${\bar K}_{\rm e-s}({\bf k}_{\bf F},{\bf p}={\bf k}_{\bf F},\omega)$ on the constant energy contour ${\bf k}_{\rm F}$, while the other broad peak with the overlap of the weights of the quasiparticle spectral function with kernel function ${\bar K}_{\rm e-s}({\bf k}_{\bf F},{\bf p}={\bf k}_{\rm BS},\omega)$ on the constant energy contour ${\bf k}_{\rm BS}$. The qualitative agreement between the present theoretical results based on the kinetic-energy driven SC mechanism and the deduced data from the ARPES measurements \cite{Bok10,Bok16,Yun11,Schachinger08} therefore show that superconductivity is indeed driven by the coupling between the electrons and spin excitations. Finally, we emphasize that as in our previous studies \cite{Gao18a,Gao18b,Feng16}, the commonly used parameters in this paper are chosen as $t/J=2.5$ and $t'/t=0.3$ for a qualitative discussion. Although the values of $J$, $t$, and $t'$ are believed to vary somewhat from compound to compound \cite{Damascelli03}, we believe that the essential feature of the coupling strength of the electrons to spin excitations in cuprate superconductors does not change with the variation of these parameters.

\section{Conclusions}\label{conclusions}

Within the framework of the kinetic-energy-driven SC mechanism, we have studied how the coupling strength of the electrons to spin excitations in cuprate superconductors evolves with momentum, energy, and doping. The normal self-energy in the particle-hole channel and pairing self-energy in the particle-pariticle channel generated by the interaction between electrons by the exchange of spin excitation have been employed to extract the coupling strengths of the electrons to spin excitations in the particle-hole and particle-particle channels, respectively. Our results show that below $T_{\rm c}$, both the coupling strengths in the particle-hole and particle-particle channels at around the antinodes consist of two peaks, with a sharp low-energy peak located at around 5 meV in the optimally doped regime, and a broad peak centered at around 40 meV, followed by a flat featureless region extending to high energy. In particular, we show that although the weights of two peaks are suppressed as the temperature is raised, this two-peak structure in the coupling strength in the particle-hole channel can persist into the normal-state. However, as a consequence of the d-wave type symmetry of the SC gap, the coupling strength in the particle-particle channel vanishes at the nodes. Moreover, our results indicate that these coupling strengths are doping dependent, where the positions of the peaks in the coupling strengths in the underdoped regime shift towards to higher energies with the increase of doping. On the other hand, the coupling strengths that we find have a striking momentum dependence, with the positions of the peaks in the coupling strengths that move to lower energies from the antinode to the node on EFS, while the weights of the peaks decrease smoothly with the move of the momentum from the antinode to the hot spot. Our theory also predicts that both the coupling strengths in the particle-hole and particle-particle channels fade away from the hot spots on EFS, which should be verified by future experiments.

\section*{Acknowledgements}

The authors would like to thank Professor Xingjiang Zhou and Dr. Deheng Gao for helpful discussions.

\section*{Disclosure statement}

No potential conflict of interest was reported by the authors.

\section*{Funding}

This work was supported by the National Key Research and Development Program of China under Grant No. 2016YFA0300304, and the National Natural Science Foundation of China under Grant Nos. 11574032 and 11734002.

\end{document}